\documentclass[aps,prd, nofootinbib, preprintnumbers, superscriptaddress, reprint]{revtex4-2}
\pdfoutput=1
\usepackage{amssymb, amsmath}
\usepackage{graphicx}
\usepackage{wrapfig}
\usepackage[normalem]{ulem} 
\usepackage[colorlinks,linktocpage]{hyperref}
\usepackage[pdftex]{pict2e}
\hypersetup{
	colorlinks=true,
	citecolor=blue
}
\usepackage{xcolor}
\newcommand{\veps}{\varepsilon}
\newcommand{\vf}{\varphi}
\newcommand{\prt}{\partial}
\newcommand{\const}{\operatorname{const}}

\newcommand{\vaa}[1]{\textcolor{red}{#1}}

\newcommand{\mbs}[1]{\boldsymbol{#1}}
\begin{document}
\preprint{INR-TH-2023-008}
\title{Analytic description of monodromy oscillons}
	\author{D.G.~Levkov}
	\email{levkov@ms2.inr.ac.ru}
	\affiliation{Institute for Nuclear Research of the
		Russian Academy of Sciences, Moscow 117312, Russia}
	\affiliation{Institute for Theoretical and Mathematical
		Physics, MSU, Moscow 119991, Russia}
	\author{V.E.~Maslov}
	\email{vasilevgmaslov@ms2.inr.ac.ru}	
	\affiliation{Institute for Nuclear Research of the
		Russian Academy of Sciences, Moscow 117312, Russia}
	\affiliation{Institute for Theoretical and Mathematical
		Physics, MSU, Moscow 119991, Russia}
	\affiliation{Department of Particle Physics and Cosmology, Faculty of Physics, MSU, Moscow 119991, Russia}
	\begin{abstract}
            We develop precise analytic description of oscillons~--- long-lived
            quasiperiodic field lumps~--- in scalar field theories
            with nearly quadratic potentials, e.g.\ the monodromy
            potential. Such oscillons are essentially nonperturbative
            due to large amplitudes, and they achieve extreme longevities. Our
            method is based on a consistent expansion in the anharmonicity
            of the potential at strong fields, which is made accurate by introducing a field-dependent ``running mass.'' At 
            every order, we compute effective action for the oscillon
            profile and other parameters. Comparison with explicit
            numerical simulations in (3+1)-dimensional monodromy model
            shows that our method is significantly more precise than
            other analytic approaches. 
	\end{abstract}
	\maketitle
	\section{Introduction}
	Oscillons \cite{Kudryavtsev:1975dj, Bogolyubsky:1976nx,
     Gleiser:1993pt, Kolb:1993hw} are long-lived pulsating field configurations that emerge in many
     classical theories, specifically, in models of a scalar field $\varphi(t,\mbs{x})$
     \cite{Salmi:2012ta, Amin:2013ika, Sakstein:2018pfd, Dorey:2019uap,
     Olle:2020qqy, Mendonca:2022fje, vanDissel:2023zva}. 
     With time, these objects demise by emitting radiation. Nevertheless, their existence may affect a wide range of
     cosmological phenomena~\cite{Lozanov:2014zfa, Liu:2017hua, Cotner:2019ykd,
     Lozanov:2023aez}, from inflationary preheating
     \cite{Amin:2011hj, Hong:2017ooe, Mahbub:2023faw,
     Aurrekoetxea:2023jwd} to phase transitions
     \cite{Copeland:1995fq,Dymnikova:2000dy, Gleiser:2010qt} and
     generation of axion dark matter \cite{Kolb:1993hw,
     Vaquero:2018tib, Buschmann:2019icd}.

\begin{figure}[b]
		\centering
		\unitlength=1mm
		\begin{picture}(80,55)
			\put(0,0){\includegraphics{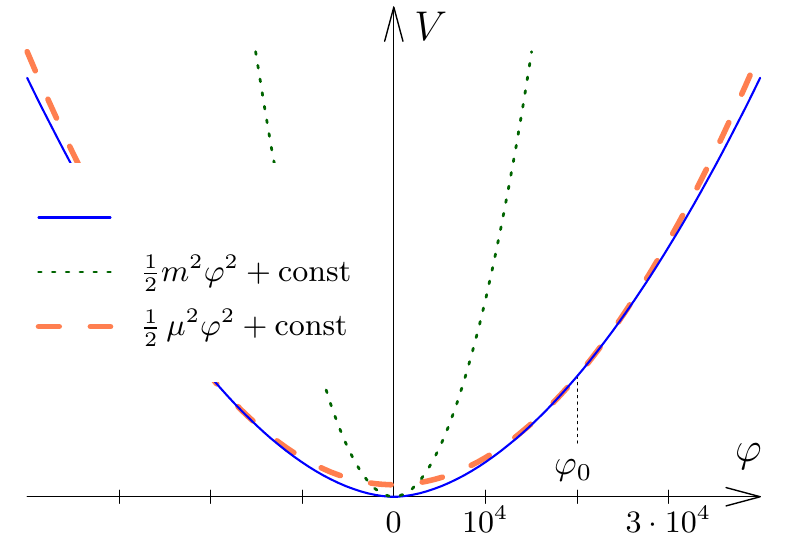}}
			\put(14.5,32){Eq.~\eqref{monodromy_welcome}}
		\end{picture}
		\caption{The potential \eqref{monodromy_welcome}
                  (solid line) approximated by the leading term of its
                  perturbative expansion around the vacuum (dotted)
                  and the first term in Eq.~\eqref{V_lin+pert}
                  (dashed). Typical field inside oscillons is
                  indicated by $\vf \sim \vf_0$. Here we use $p = 0.9$ and~$\vf_0 = 2 \cdot 10^4$.}
		\label{fig:monodromy_mu}
	\end{figure}

     In some models oscillons live exceptionally
     long~\cite{Olle:2019kbo, Olle:2020qqy, Cyncynates:2021rtf}. An 
    important example is a scalar field theory with the monodromy
    potential \cite{Amin:2011hj, Olle:2019kbo, Sang:2019ndv}
	\begin{equation}
	\label{monodromy_welcome}
	V(\varphi) = \frac{1}{2p} (1 + \varphi^2)^p\,, \qquad
         p \lesssim 1 \,,
	\end{equation}
   see Fig.~\ref{fig:monodromy_mu}. Hereafter we use
   dimensionless units\footnote{Physical dimensions can be restored by
   rescaling $\vf \to \vf/F$, ${t \to mt}$, and $\mbs{x} \to m
   \mbs{x}$, which gives $V = m^2 F^2 (1 + \vf^2/F^2)^p/2p$.} with
   particle mass $m=1$.

   Oscillons in the model~\eqref{monodromy_welcome} exist and last for
   up to $10^{14}$ periods \cite{Olle:2020qqy}. Their lifetimes are
   considerably larger at  $p \approx 1$ when the potential is almost 
   quadratic. This last property is generic: oscillons with extreme
   longevities are expected to appear in models with suppressed
   interactions~\cite{Olle:2020qqy,Levkov:2022egq}. 
   
   Monodromy oscillons significantly alter some cosmological
     scenarios described by the model~(\ref{monodromy_welcome}). They
     may  form and produce gravitational waves~\cite{Amin:2011hj, 
       Zhou:2013tsa, Sang:2019ndv} at the reheating stage of the 
     monodromy inflation~\cite{Silverstein:2008sg, McAllister:2008hb,
       Cicoli:2023opf} where~$\varphi$ represents inflaton
     and~$p=1/2$. At~${p  \approx 1}$  long-lived monodromy oscillons may
     impact cosmological evolution and even form a (part of)
     scalar field dark  matter~\cite{Olle:2019kbo, Olle:2020qqy}.  

	The aim of this paper is to develop a consistent, practical and precise analytic approach to oscillons in theories with nearly quadratic potentials. It would be logical to organize such a technique as an asymptotic expansion in the inverse oscillon lifetime~$\tau^{-1}  \ll  m$. However, the parameter $\tau$ is not built into the field theory Lagrangian and cannot be used directly. That is why all practical descriptions of oscillons use other~--- convenient~--- expansion parameters.
		
	Presently, there are two\footnote{Some other witty tricks
        \cite{Amin:2010jq,Manton:2023mdr} have limited region of
        validity and also do not apply in the
        model~\eqref{monodromy_welcome}.} general analytic approaches
        to oscillons, and both are useless in the
        model~\eqref{monodromy_welcome}.  The first hope is a
        perturbative expansion~\cite{Kosevich1975, Dashen:1975hd,
          Segur:1987mg, Fodor:2008es, Fodor:2019ftc}, since weakness
        of interactions may suppress wave emission and guarantee
        longevity. For the potential~\eqref{monodromy_welcome}, this
        is equivalent to expansion in small field amplitudes
        $|\varphi| \ll 1$ which~--- alas --- does not work  inside the
        oscillons because their fields are strong,~$\vf \sim \vf_0 \gg
        1$. We demonstrate this in Fig.~\ref{fig:monodromy_mu} by
        comparing at~${|\varphi| \gg 1}$ the potential with its
        quadratic part $m^2 \vf^2/2$ (solid and dotted lines).
	
	The second ``Effective Field Theory'' (EFT)
        approach~\cite{Levkov:2022egq} is based on the observation
        \cite{Gleiser:2004an} that spatial sizes of the longest-living
        oscillons are large,~${R \gg m^{-1}}$. This is true, in
        particular, for oscillons in the
        model~\eqref{monodromy_welcome}. 
	It is natural to expect that such large objects evolve slowly
        at time scales~$t \sim R$ considerably exceeding their
        oscillation periods~$T \sim m^{-1}$.  
	Then their stability can be related to the  conservation of
        adiabatic invariant~$N$ \cite{
          Kasuya:2002zs,Kawasaki:2015vga}. 
	Nevertheless, direct asymptotic expansion in~$R^{-1}$
        \cite{Levkov:2022egq} is of a limited practical use because it
        exploits general nonlinear spatially homogeneous solutions
        which cannot be obtained analytically in the
        model~\eqref{monodromy_welcome}.

	In this paper we merge the above two approaches together into
        a general, simple and precise analytic technique applicable in
        the model~\eqref{monodromy_welcome}. On the one hand, we
        remedy the perturbative method by noting that the
        potential~\eqref{monodromy_welcome} with $p$ somewhat
          below~1 can be approximated by a parabola $\mu^2
          \varphi^2/2$ in every vast  
          region~$\vf \sim O(\vf_0)$, where~$\varphi_0$ is arbitrarily
          large. The latter parabolas, however, essentially differ from the leading
          Taylor term of the potential near the vacuum, as their
          curvatures~$\mu^2$ change with~$\vf_0$~--- cf.\ the dashed, solid, and
          dotted  lines in Fig.~\ref{fig:monodromy_mu}. This means
        that the scalar particles interact weakly even in the
        strong-field regions, but their ``mass''~$\mu$ slowly depends
        on~$\varphi$. We therefore write 
	\begin{equation}
		\label{V_lin+pert}
		V(\vf) = 
		\mu^2 \vf^2/2 + \delta V(\vf)\;, \quad \delta V \equiv 
			V(\varphi)- \mu^2 \vf^2/2\,,
	\end{equation}
	where $\mu \neq m$, and perform expansion\footnote{In
        Eq.~\eqref{V_lin+pert} we ignore constant part of the
        potential which does not affect the field equation.} in
        $\delta V$. Since Eq.~\eqref{V_lin+pert} is a trivial
        identity, the result does not depend on~$\mu$ which is made
        field-dependent in the end of the calculation. We will see
        that this trick with field-dependent~$\mu$ radically increases
        precision, working in a similar way to scale-dependent
        renormalized constants in QFT. On the other hand, weak
        interactions imply large sizes of bound states~---
        oscillons~--- so we perform further expansion in the inverse
        oscillon radius $R^{-1}$. The accuracy of the overall 
          method does not deteriorate even at extremely large
          oscillon amplitudes, although this technique is less
          effective for essentially nonlinear potentials with small~$p$.
	
	To sum up, our expansion parameters are ${(\prt_i \vf)^2 \ll \mu^2 \vf^2}$ and $\delta V \ll \mu^2 \vf^2$. At every order, we analytically obtain effective action for the oscillon profiles, find expressions for their adiabatic charges $N$ and energies $E$. At the roughest level, our method is close to the single-frequency approximation, a popular and practical heuristic technique~\cite{Zhang:2020bec,Cyncynates:2021rtf,vanDissel:2023zva,Eby:2014fya}. We provide a recipe for computing corrections thus upgrading this technique to a consistent asymptotic expansion.
	
	The paper is organized as follows. We start in Sec.~\ref{sec:oscillator} by illustrating the new method in a toy mechanical model. Then we apply it to field-theoretical oscillons in Sec.~\ref{sec:EFT} and confirm its predictions with exact numerical simulations in Sec.~\ref{sec:Comparison}. Corrections are considered in Sec.~\ref{sec:Corrections}. Section~\ref{sec:Discussion} contains discussion and comparison to other analytic approaches.
	
	\section{Toy model: a mechanical oscillator}
	\label{sec:oscillator}
	For a start, let us illustrate the key trick of this paper in the mechanical model of nonlinear oscillator with the potential~\eqref{monodromy_welcome}. Its coordinate $\vf = \vf(t)$ satisfies the equation
	\begin{equation}
		\label{nl_osc_eq}
		\partial^2_t \varphi = -  V'(\vf)
        \equiv -\varphi \left(1 + \varphi^2\right)^{-\veps}, 
        \quad \veps \equiv 1-p,
	\end{equation}
	which is not exactly solvable; here the prime denotes $\vf$ derivative.

    At $\veps \ll 1$ and $p \approx 1$, however, the nonlinearities are small, so we can approximately integrate the equation as follows.
	We add the auxiliary quadratic term $\mu^2 \vf^2/2$ to the potential and subtract it back,
		\begin{equation}
		\label{osc_eq_f0}
		\partial^2_t \varphi = -\mu^2 \varphi
                - \delta V'(\vf)\,,
                \quad \quad \delta V \equiv V - \mu^2 \vf^2/2,
                \end{equation}
	cf.\ Eq.~\eqref{V_lin+pert}.
	The nonlinear force $\delta V'$ is weak if $\mu^2$ is close to the curvature of the potential.
	We achieve this by choosing
	\begin{equation}
		\label{mu_0_def}
		\mu^2 = V'(\vf_0)/\vf_0 \equiv \left(1 + \vf^2_0\right)^{-\veps},
	\end{equation}
	where the ``renormalization scale'' $\vf_0$ will be tuned soon to the typical oscillator amplitude. The proper choice \eqref{mu_0_def} allows us to develop a consistent expansion in small $\delta V$.
	
To the zeroth order in the last term, Eq.~\eqref{osc_eq_f0} describes linear oscillations with frequency $\mu$. It can be solved by performing a canonical transformation from the coordinate $\vf$ and momentum~$\pi_\vf \equiv \prt_t \vf$ to the action-angle variables,
	\begin{equation}
		\label{lin_I_theta}
		\varphi = \sqrt{2I/\mu}\,  \cos \theta, \qquad
                \pi_\varphi =
                -\sqrt{2I\mu} \sin \, \theta\,.
        \end{equation}
  	Here $I(t)$ and $\theta(t)$
	characterize the amplitude and phase of the oscillations, respectively. Namely, the evolution without the $\delta V$ term would be $I = \const$ and~$\theta = \mu t$.
	
	The next step is to include nonlinear corrections in~$\delta V$.
	It is convenient to do that on the level of classical action,
	\begin{equation}
		\label{osc_action}
		\mathcal{S} = \int dt \left[\pi_\vf \prt_t \vf - H \right] = \int dt \left[I \prt_t \theta - H(I, \theta) \right],
			\end{equation}
			where the Hamiltonian is
	\begin{equation}
		\label{osc_hamilt}
	  H = \frac12 \pi^2_\vf + V(\vf) = \mu I + \delta V(I, \theta)
	\end{equation}
	and we performed the transformation~\eqref{lin_I_theta} in the second equalities. 
	
	It is worth stressing that $I$ and $\theta$ are not the true action--angle variables of the full nonlinear oscillator.
	Hence, the perturbation $\delta V$ may cause $I$ and $\prt_t \theta$ to drift slowly, and also equips them with tiny oscillating corrections. Below we consider stationary solutions on long timescales. In this case the integral~\eqref{osc_action} averages the perturbations over many periods. Since $\theta$ evolves linearly in the zeroth--order approximation, we can perform the averaging by integrating\footnote{This and other $\theta$ averages can be recast as contour integrals ${\langle f \rangle =  \oint  f(z) dz /(2 \pi i z)}$ over the unit circle ${z = \mathrm{e}^{2i\theta}}$, $|z|=1$.} over it,
        \begin{equation}
		\label{average_theta}
		\delta V \to \langle \delta V \rangle = \int\limits_0^{2\pi}
                \frac{ d\theta }{2\pi} \, \delta V(\theta)  
		= \frac1{2p} \left(\mathcal{A}_p (\varsigma) - p\mu I \right),
  \end{equation}
  	where $\varsigma = 2I/\mu$,
	\begin{multline}
     \label{Av_p_sigma}
		  \mathcal{A}_p (\varsigma)   \equiv \left\langle
                  \left(1 + \varsigma \cos^2
                  \theta\right)^p\right\rangle \\
                  = (1 + \varsigma) ^{p/2} P_p\left(\frac{1 +
                      \varsigma/2}{\sqrt{1 + \varsigma} }
                  \right),
	\end{multline}
     and $P_p(x) \equiv {}_2 F_1 [-p, \, p+1; \, 1;\, (1-x)/2]$ is the Legendre function.
     
	Together, Eqs.~\eqref{osc_action}, \eqref{osc_hamilt}, and~\eqref{average_theta} form an effective action describing long-term behavior of the monodromy oscillator. We obtained it to the leading order in $\delta V$.
	
	Now, we make a crucial step: choose the parameter~$\mu$ and its arbitrary ``scale'' $\vf_0$ in Eq.~\eqref{mu_0_def}. The approximate action~\eqref{osc_action}, \eqref{osc_hamilt}, \eqref{average_theta} depends on $\vf_0$, but very weakly: its variation with respect to that parameter is~$O(\delta V^2) \sim O(\veps^2)$. Indeed, $\mu$ cancels itself in the exact action~\eqref{osc_action} and resurfaces in the approximate results only because we ignored $\mu$-dependent corrections. If we continued computations to the $n$-th order in~$\delta V$, the sensitivity of the effective action to $\vf_0$ and $\mu$ would be~$O\left(\veps^{n+1}\right)$. Thus, we are free to choose $\vf_0$ in any reasonable way that decreases $\delta V$ and increases the precision of the expansion.
	Specifically, we adjust $\vf_0$ to the field amplitude\footnote{Alternatively, one can find $\vf_0(I)$ from the equation ${\vf^2_0 = 2I/\mu}$, where $\mu$ is still given by Eq.~\eqref{mu_0_def}. This method gives slightly better results at extremely large amplitudes when $\mu$ is significantly smaller than $1$.\label{fn:smart_mu}} in Eq.~\eqref{lin_I_theta},
	\begin{equation}
		\label{phi0_choose}
		\vf_0 = \sqrt{2I}.
	\end{equation}
	This makes $\mu = \mu(I)$ depend on the action variable.
	
	Equations for the long-term motions of the monodromy oscillator are obtained by varying the effective action~\eqref{osc_action}, \eqref{osc_hamilt}, \eqref{average_theta} with respect to $I(t)$ and $\theta(t)$. We get~$\theta = \omega t$ and
	\begin{equation}
		\label{osc_frequency}
		\omega = \mu + \left(
                  \partial_\varsigma\mathcal{A}_p/\mu^2 p  - 1/2
                  \right) (\mu - I \partial_I \mu) \,,
	\end{equation}
	where $\varsigma = 2I / \mu$ and
    $\partial_I \mu = -\veps (1+2I)^{-\veps/2-1}$. Equation~\eqref{osc_frequency} gives oscillation frequency $\omega = \omega(I)$ as a function of the action variable. 

\begin{figure}
	  \centering
          \unitlength=1mm
          \begin{picture}(86,61)
	    \put(0,0){\includegraphics{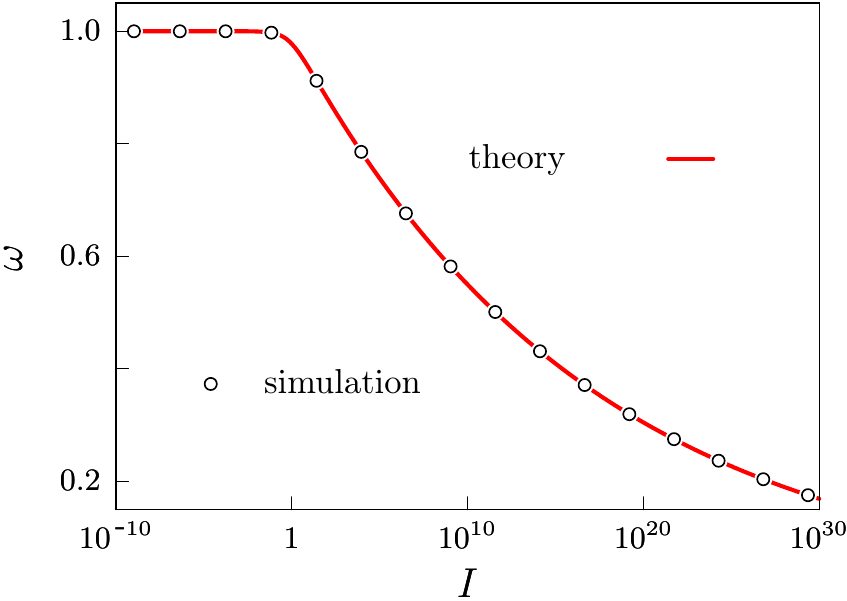}}
            \put(59,44){\eqref{osc_frequency}}
          \end{picture}
		\caption{Frequency $\omega$ of the monodromy oscillator as a function of the action variable $I$; we use $p=0.95$. Solid line and circles show the theory~\eqref{osc_frequency} and exact numerical results, respectively.}
		\label{fig:oscillator}
	\end{figure}

	To test the method, we compare Eq.~\eqref{osc_frequency} with the exact frequency $\omega(I)$ of the monodromy oscillator at ${p=0.95}$; see Fig.~\ref{fig:oscillator}. The latter is computed by numerically evolving Eq.~\eqref{nl_osc_eq} and then extracting the oscillation period $T = 2 \pi/\omega$ and the action variable $I = \oint \pi_\varphi\, d\vf/2\pi$. Figure~\ref{fig:oscillator} shows that the theory has a remarkable relative precision $\Delta \omega/\omega \sim 10^{-4}$ which remains stable even in the case of exceptionally large amplitudes $I \sim 10^{30}$.
			
	The secret of high accuracy is hidden in two tricks. First, we correctly selected the values of artificial parameters $\varphi_0$ and $\mu(\vf_0)$, and even allowed them to ``run''~--- depend on~$I$~--- in the end. This approach is similar to running of the renormalized constants in QFT.
	Second, despite working at small $\veps \equiv 1- p$, we did \textit{not} directly expand $\delta V$ in this parameter: the latter expansion actually goes in $\veps \ln |\vf|$ and breaks at large amplitudes~${|\vf| \gtrsim \mathrm{e}^{1/2\veps}}$.
	
	It is worth noting that corrections to the effective action~\eqref{osc_action}, \eqref{osc_hamilt}, \eqref{average_theta} can be computed in a straightforward manner, by taking higher-order $\delta V$ contributions into account. We will work them out in Sec.~\ref{sec:Corrections} when effective action in field theory is introduced.

	\section{Effective field theory for oscillons}
	\label{sec:EFT}
	Let us turn to oscillons --- long-lived solutions of the field equation 
	\begin{equation}		
		\label{field_eq}
		\partial^2_t \varphi - \Delta \varphi = - V'(\varphi)
	\end{equation}
        with the monodromy potential~\eqref{monodromy_welcome}. For definiteness,  we will work in $3+1$ dimensions: generalizations to other cases are straightforward.
        We perform the trick~\eqref{V_lin+pert}: extract the quadratic part of the potential~$\mu^2 \vf^2/2$ and then work order-by-order in the remaining part $\delta V$.
Now, the equation includes a spatial derivative $\Delta \vf$ which also can be  treated perturbatively. Indeed,~$\Delta \vf$ and $\delta V'$ are comparable inside the oscillons because these objects are held by weak attractive self-force compensating repulsive contributions of the derivatives (``quantum pressure''). As a consequence, the oscillon sizes are parametrically large:~$R^{-2} \sim O(\delta V) \sim O(\veps)$.
	
To the zeroth order in $\delta V$, we have the same linear oscillator $\prt_t^2 \vf = -\mu^2 \vf$ as in the previous section, where~$\mu(\vf_0)$ again estimates local curvature of the potential via Eq.~\eqref{mu_0_def}.

The next step is to perform the transformation~\eqref{lin_I_theta} to the action-angle variables, which are now the fields $I(t, \mbs{x})$ and $\theta(t, \mbs{x})$. The classical
        action takes the form
        \begin{align}
          \notag 
	\mathcal{S} &= \int d^4 x \left[ \pi_\vf \partial_t \vf  -
            \pi_\vf^2/2  - (\partial_i \varphi)^2/2 -  V(\varphi) \right] \\ 
        \label{field_action}
        & = 
        \int dt \, d^3 \mbs{x}\, \left[ I \partial_t \theta - \mu I-
          (\partial_i \vf)^2/2 - \delta V\right]\,.
\end{align}
As before, we average the perturbations --- the two last terms in Eq.~\eqref{field_action}~--- over period.
Expression for $\langle \delta V \rangle$ is already given in Eq.~\eqref{average_theta}. To process the term with spatial derivatives, we substitute Eq.~\eqref{lin_I_theta} into $(\prt_i \vf)^2$, move slowly varying $\prt_i I$ and $\prt_i \theta$ out of the average, and then integrate the coefficients in front of them over $\theta$, cf.\ Eq.~\eqref{average_theta}. This gives
	\begin{equation}
		\label{aver_spatial}
          \left\langle (\prt_i \vf)^2 \right\rangle =
          \frac{(\prt_i I)^2}{4I\mu} + \frac{I}{\mu} \, (\prt_i \theta)^2\,,
	\end{equation}
	where the cross-term $\prt_i I  \prt_i \theta$ vanishes due to  time-reflection symmetry $\theta \to -\theta$. 
	
	Substituting Eqs.~\eqref{average_theta} and~\eqref{aver_spatial} into Eq.~\eqref{field_action}, we arrive at the leading-order effective action for oscillons,\footnote{One may pack real $I$ and $\theta$ into one complex field ${\psi(t, \mbs{x}) = \sqrt{I}\mathrm{e}^{-i\theta}}$, see Ref.~\cite{Levkov:2022egq}. This turns the effective theory~\eqref{field_eff_action} into a nonlinear Schr\"odinger model with global symmetry $\psi \to \psi \,\mathrm{e}^{-i\alpha}$.}
	\begin{align}
          \notag
	 \mathcal{S}_{\textrm{eff}} = \int dt \, d^3 &\mbs{x} \,
         \left[I \partial_t \theta - \mu I \right.\\ 	\label{field_eff_action}
           &\left. 
           \!\! - \frac{(\prt_i I)^2}{8I\mu}
           - \frac{I(\prt_i \theta)^2}{2\mu}
           -\frac{\mathcal{A}_p(\varsigma)}{2p}
           + \frac{\mu I}{2}\right],
	\end{align}
	where $\varsigma = 2I/\mu$ and the function $\mathcal{A}_p (\varsigma)$ is defined in Eq.~\eqref{Av_p_sigma}.
	
	The final --- important --- step is to make the ``effective mass'' $\mu = \mu(I)$ and its ``scale'' $\vf_0 = \vf_0(I)$ depend on the oscillation amplitude\footnote{One can show that the effective action~\eqref{field_eff_action} is insensitive to $\vf_0$ up to $O(\veps^2)$ corrections, where $\delta V \sim (\partial_i I)^2/I \sim O(\veps)$.} via Eqs.~\eqref{mu_0_def} and \eqref{phi0_choose}. Like in Sec.~\ref{sec:oscillator}, this tunes $\mu$ to the second derivative of the potential at any $I$.
	
	One can see straight away that the effective model~\eqref{field_eff_action} is invariant under the global phase shifts $\theta \to \theta + \alpha$. This implies conservation of the global charge
	\begin{equation}
		\label{N_charge}
		N = \int d^3\mbs{x} \; I(t,\, \mbs{x})
	\end{equation}
	representing the adiabatic invariant of Refs.~\cite{Kasuya:2002zs,Kawasaki:2015vga}.
	
	From the viewpoint of the effective theory, oscillons are nontopological solitons \cite{Coleman:1985ki, Nugaev:2019vru} minimizing the energy $E$ at a fixed charge $N$. Their profiles can be obtained by extremizing the functional $F = E - \omega N$ with Lagrange multiplier~$\omega$, or simply by substituting the stationary ansatz
	\begin{equation}
		\label{stat_ansatz}
		I(t, \mbs{x}) = \psi^2(\mbs{x}), \qquad \theta(t, \mbs{x}) = \omega t,
	\end{equation}
	into the effective field equations. Either way, the resulting equation for $\psi(\mbs{x})$ is
	\begin{multline}
		\label{profile_equation}
		  \omega \psi = \mu\psi
                  -\frac{\Delta \psi}{2\mu} + \psi (\prt_i \psi)^2 \,
                  \frac{\prt_I\mu}{2\mu^2}  \\ +
                  \left(\partial_\varsigma \mathcal{A}_p/ \mu^2
                    p  - 1/2 \right)
                  (\mu - \psi^2
                  \partial_I \mu)  \psi , 
	\end{multline}
	cf. Eq.~\eqref{osc_frequency} and recall that $\mu = \mu(I)$ is different from the ``bare'' field mass $m=1$.	
	\begin{figure}[h]
		\centering
		\includegraphics{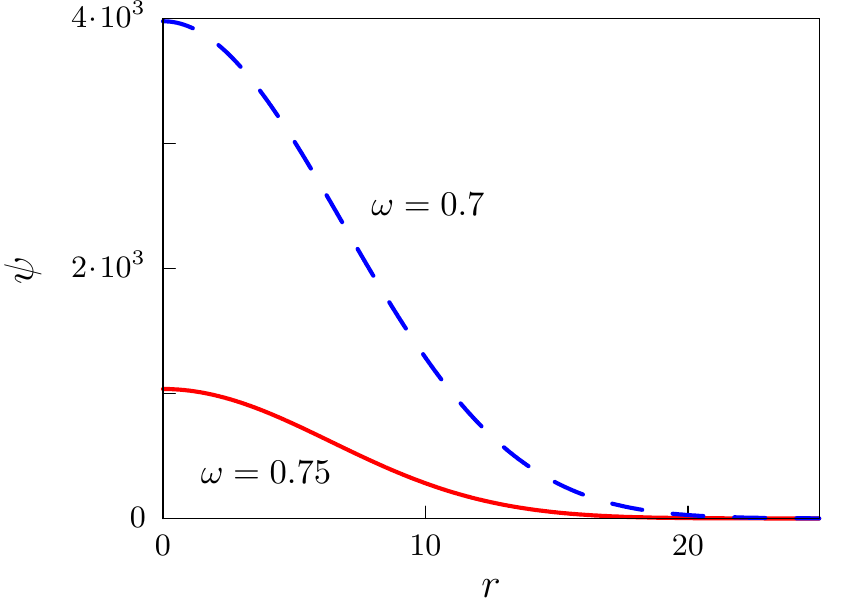}
		\caption{Profiles $\psi(r)$ of the monodromy oscillons at two 
                  frequencies and~$p = 0.95$.}
		\label{fig:psi_profiles}
	\end{figure}
	
	We solve Eq.~\eqref{profile_equation} in spherical symmetry 
	using the shooting method. Figure~\ref{fig:psi_profiles} shows the solutions $\psi = \psi(r)$ at two values of $\omega$. They characterize the oscillon amplitudes which are large and rapidly grow as the frequency~$\omega$ decreases.
	
	\section{Comparison with numerical simulations}
	\label{sec:Comparison}
	Let us compare the predictions of our effective field theory (EFT) with exact oscillons. We obtain the latter by numerically evolving the full equation~\eqref{field_eq} for the spherically symmetric field $\vf(t,r)$. We start these simulations from the EFT oscillons, i.e.\ $\vf(0,r)$ given by the profiles~$\psi(r)$ and Eqs.~\eqref{stat_ansatz}, \eqref{lin_I_theta}. Practice shows that such initial data settle down to true non-excited oscillons much quicker than generic initial conditions. Details of our numerical method are presented in Ref.~\cite{Levkov:2022egq}.
	
\begin{figure}
		\centering
		\includegraphics{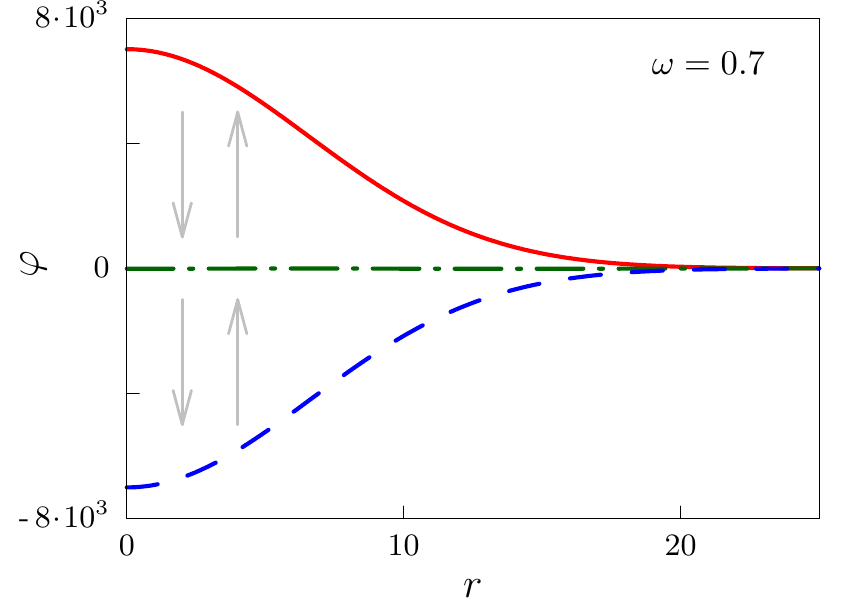}
		\caption{Numerical evolution of the monodromy oscillon during one period. We use $\omega = 0.7$ and $p=0.95$.}
		\label{fig:phi_evolve_period}
	\end{figure}

	Typical evolution of the oscillon field during one period is visualized in Fig.~\ref{fig:phi_evolve_period}, where solid and dashed lines correspond to oscillation maximum and minimum, respectively. In the exact model, outgoing radiation carries away the energy and makes the oscillon amplitude decrease, but that process is extremely slow and cannot be seen in Fig.~\ref{fig:phi_evolve_period}.
	
	Once the exact oscillon is formed, we measure its period~${T = 2\pi/\omega}$ as the time interval between the consecutive maxima of the field at the center  $\vf(t, 0)$. The ``exact'' value of the adiabatic invariant  $N$ is given by the standard formula \cite{Kasuya:2002zs,Kawasaki:2015vga},
	\begin{equation}
		\label{N_adia}
		N = \frac{1}{2\pi}\int d^3 \mbs{x} \oint \pi_\vf \,d\vf = \int d^3 \mbs{x} \int\limits_t^{t+T} \frac{dt}{2\pi} \,(\prt_t\vf)^2\!.
	\end{equation}
	Finally, the oscillon energy $E$ is
	\begin{equation}
		E = \int d^3 \mbs{x} \left[(\prt_t \vf)^2/2 + (\prt_i \vf)^2/2 + V(\vf)\right].
	\end{equation}
	To increase precision, we average all ``exact'' quantities over several periods.
	
\begin{figure}
		\centering
		\includegraphics{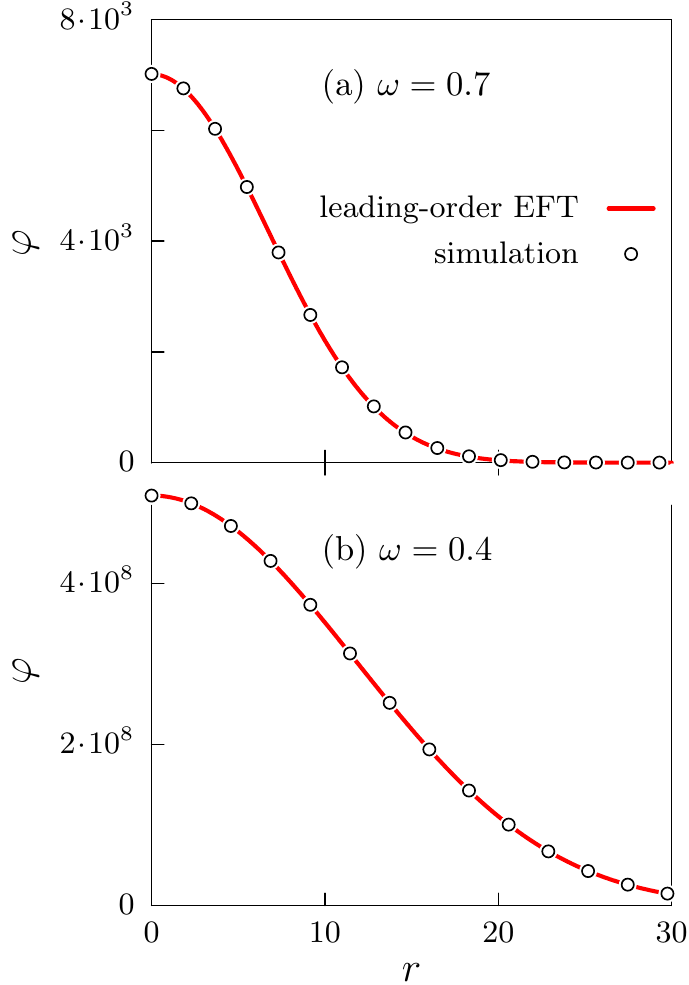} 
		\caption{Fields $\vf(t_{\max}, r)$ of the monodromy oscillons at times~$t_{\max}$ of oscillation maxima. We use $p = 0.95$ and frequencies (a)~${\omega=0.7}$, (b) $\omega=0.4$. Solid line is the EFT prediction, while the circles show results of full numerical simulations.}
		\label{fig:profiles_compare}
	\end{figure}

\begin{figure}
		\centering
		\includegraphics{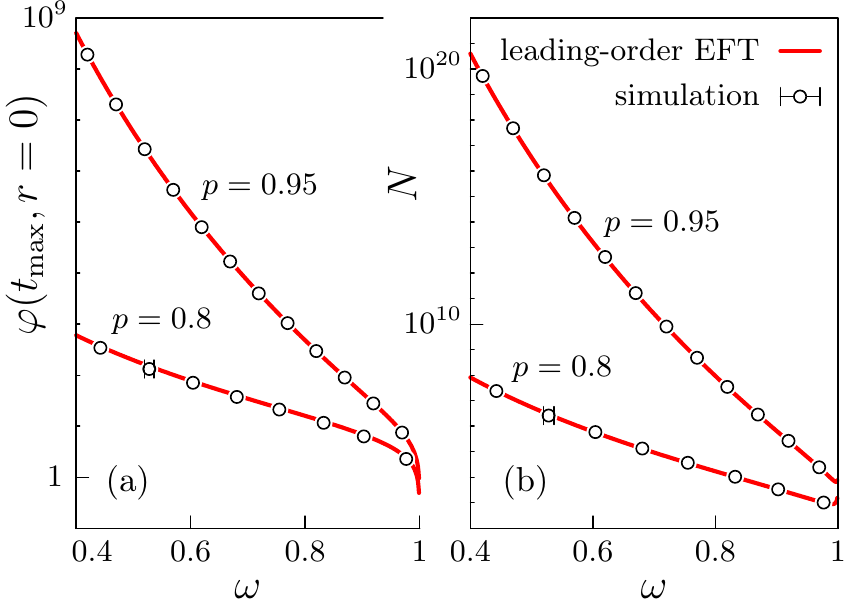}
		\caption{(a) Amplitudes $\vf(t_{\max},\, r=0)$ and
                    (b) adiabatic invariants $N(\omega)$ of the monodromy
                  oscillons at $p=0.8$ and $p=0.95$. Solid
                  lines show the predictions of the leading-order
                  EFT. At two values of~$p$, they reproduce the exact numerical results
                  (circles) with relative precisions $\Delta N / N
                  \sim 0.2$ and~$0.06$, respectively. The latter accuracies  are almost
                    independent of the oscillon amplitude.}
		\label{fig:Om_N}
	\end{figure}

	In Fig.~\ref{fig:profiles_compare} we compare the fields $\vf(t, \mbs{x})$ of numerical and EFT oscillons with the same frequency~$\omega$ at the moments~$t_{\max}$ of oscillation maxima, $\prt_t \vf(t_{\max},\, 0) = 0$. 
	The EFT prediction (lines) is obtained by substituting the  profile $I = \psi^2(r)$ into Eq.~\eqref{lin_I_theta} at $\theta = 0$. It agrees well with the numerical results (circles) at two essentially different values of~$\omega$. Note that our theory remains precise even for the $\omega = 0.4$ oscillon  that has extremely large amplitude. We further confirm this in Fig.~\ref{fig:Om_N}a showing the maximal fields $\varphi(t_{\max},\,r=0)$ of exact and EFT oscillons as functions of frequency.
	
	Good agreement is also observed for the integral quantities, say, the oscillon charge\footnote{Graphs for $E(\omega)$ and $N(\omega)$ have similar shapes due to the relation~$dE/dN = \omega$ holding for the EFT oscillons \cite{Levkov:2022egq}.} $N(\omega)$ given by Eq.~\eqref{N_charge} in the EFT and by Eq.~\eqref{N_adia} in full theory, cf.\ the lines and circles in Fig.~\ref{fig:Om_N}\vaa{b}. It is worth noting that oscillons with ${\omega \lesssim 0.99}$ satisfy the Vakhitov-Kolokolov criterion \cite{vk,Zakharov12,Levkov:2022egq} $dN/d \omega < 0$ which is necessary for stability. Thus, they are expected to be unstable with respect to linear perturbations only in the narrow frequency region $\omega \approx m = 1$.
	
	To sum up, our leading-order EFT with field-dependent ``mass''
        $\mu = \mu(I)$ gives accurate predictions even if the
        monodromy potential is moderately nonlinear, see the graph
        with $p = 0.8$ and $\veps = 0.2$ in
        Fig.~\ref{fig:Om_N}. Their relative error is almost
          insensitive to the field strength inside the oscillon but
          grows with nonlinearity of the potential. 
	
	\section{Higher-order corrections}
	\label{sec:Corrections}
	So far, we used the leading-order approximation: assumed that~$I$ and~$\theta$ evolve slowly while evaluating the average $\delta V \to \langle \delta V \rangle$ in Eq.~\eqref{average_theta}. Let us show that this approach can be promoted to a consistent asymptotic expansion in the nonlinearity $\delta V$ of the potential and in the inverse oscillon size $R^{-1}$. 
	
	We split $I$ and $\theta$ into EFT fields $\bar{I}$, $\bar{\theta}$
    evolving at large timescales and 
        fast-oscillating corrections~$\delta I$,~$\delta \theta$,
	\begin{equation}
		\label{fast_osc_delta}
		I = \bar{I}(t, \mbs{x}) + \delta I(t,
                \mbs{x})\,, \qquad
 	        \theta = \bar{\theta}(t, \mbs{x}) + \delta \theta(t, \mbs{x}),
	\end{equation}
	where $\langle\delta I \rangle = \langle\delta \theta \rangle = 0$. Effective action for $\bar{I}$ and $\bar{\theta}$ can be computed by integrating out $\delta I$ and $\delta \theta$ in the arbitrary slowly-changing background.
	To this end, we obtain exact equations for $I$ and $\theta$ from the action~\eqref{field_action} and subtract their time averages. We get,
	\begin{equation}
		\label{delta_eq_w_sources}
		\prt_t \,\delta I = j_\theta(I, \theta), \qquad \prt_t \, \delta \theta = -j_I(I, \theta),
	\end{equation}
        where
         \begin{align}
            \label{eq:1}
            & j_\theta = \partial_\theta \vf (\Delta \vf - \delta V')\;,
            \\
            & j_I = \partial_I \vf (\Delta \vf - \delta V')
            - \Big\langle \partial_I \vf (\Delta \vf - \delta V') \Big\rangle
          \end{align}
        are the sources depending on $\bar{I}$, $\bar{\theta}$, $\delta I$, and $\delta \theta$ via Eqs.~\eqref{lin_I_theta} and~\eqref{fast_osc_delta}.
          
    It is clear that Eqs.~\eqref{delta_eq_w_sources} can be used to express $\delta I$ and~$\delta \theta$ as functions of~$\bar{I}$ and~$\bar{\theta}$. Indeed, let us change the time variable to $\bar{\theta}$ which evolves progressively: ${\prt_t \approx (\prt_t \bar{\theta}) \, \prt_{\bar{\theta}}}$. Then, solving the equations order-by-order in small $\delta I$ and $\delta \theta$, we indeed find the fast-oscillating parts as series of functions depending on $\bar{I}$ and $\bar{\theta}$.

   Let calculate the first nontrivial (second-order) correction to the effective action considering the stationary oscillon background $\bar{I} = \psi^2(\mbs{x})$ and~${\bar{\theta} = \omega t}$ in Eq.~\eqref{stat_ansatz}.
   At this level, we ignore $\delta I$ and $\delta \theta$ in the right-hand sides of Eqs.~\eqref{delta_eq_w_sources}.  Then
     \begin{align}
          	\label{j_theta_mndr}
	j_\theta& \approx \frac{\sin 2 \bar{\theta}}{\mu} \psi^2 
        \left[-\Delta \psi / \psi - \mu^2 + (1 +
          \varsigma \cos^2 \bar{\theta} )^{-\epsilon} \right],\\ \notag
	j_I&\approx \frac{\cos 2 \bar{\theta}}{2\mu} 
        \left(\frac{\Delta \psi}{\psi} + \mu^2 \right) -
        \frac{\cos^2 \bar{\theta}}{\mu}
        (1+\varsigma \cos^2 \bar{\theta})^{-\epsilon} \\
        \label{j_I_mndr}
        & \qquad \qquad\quad \qquad\qquad + \frac{1}{2\psi^2}
        \left[\mathcal{A}_p(\varsigma) -
          \mathcal{A}_{p-1}(\varsigma)\right]\!, 
\end{align}
 where Eqs.~\eqref{lin_I_theta}, the monodromy potential~\eqref{monodromy_welcome}, \eqref{V_lin+pert}, and $\varsigma = 2\bar{I}/\mu$ were used. The solutions $\delta I(\bar{I}, \bar{\theta})$ and~$\delta \theta(\bar{I}, \bar{\theta})$ are given by the primitives of the sources~\eqref{j_theta_mndr}, \eqref{j_I_mndr} with respect to $\bar{\theta}$. We substitute them into the action~\eqref{field_action} expanded quadratically in $\delta I$ and $\delta \theta$ and average the result over $\bar{\theta}$. This gives the second-order effective action $\mathcal{S}_{\mathrm{eff}} = \mathcal{S}^{(1)}_{\mathrm{eff}} + \mathcal{S}^{(2)}_{\mathrm{eff}}$, where $ \mathcal{S}^{(1)}_{\mathrm{eff}}[\bar{I}, \bar{\theta}]$ is provided by Eq.~\eqref{field_eff_action} and the correction is
\begin{align}
	 \notag \mathcal{S}^{(2)}_{\mathrm{eff}} & =
          \frac{1}{\omega} \int dt \, d^3 \mbs{x} \; \left\langle \,  j_I(\bar{I}, \bar{\theta}) \int^{\bar{\theta}} j_\theta(\bar{I}, \bar{\theta}') d\bar{\theta}' \,\right\rangle\\ \notag
           & = \frac{1}{4\omega} \int dt \, d^3 \mbs{x} \, 
          \left\{\frac{1}{2\mu^2}(\Delta \psi + \mu^2 \psi)^2 +
            \frac{\mathcal{D}_p(\varsigma)}{p\psi^2}  
            \right. \\  & \qquad \qquad \qquad \qquad\quad
            -\left.\frac{\mathcal{B}_p(\varsigma)}{p\psi^3} \, (\Delta \psi +
            \mu^2 \psi) 
            \right\}\,,
          	\label{S_corr}
\end{align}
see Ref.~\cite{Levkov:2022egq} for general and detailed discussion.  
In Eq.~\eqref{S_corr} we introduced the form factors
\begin{align}
	\mathcal{B}_p & = (p+1) \,
  \mathcal{A}_{p+1}(\varsigma)+ p\,(\varsigma/2+1) \,
  \mathcal{A}_{p-1}(\varsigma) \notag \\
  & \qquad \qquad \qquad \; - (1+2p+(p+1)\,\varsigma/2)
  \,\mathcal{A}_{p}(\varsigma)  \,,
 	\\
 	\mathcal{D}_p &=
        \mathcal{A}_{2p}(\varsigma) - \mathcal{A}^2_{p}(\varsigma) 
        - \mathcal{A}_{2p-1}(\varsigma) +
        \mathcal{A}_{p}(\varsigma)\mathcal{A}_{p-1}(\varsigma)
\end{align}
in front of the terms with different numbers of derivatives.

The second-order effective theory remains invariant under the shifts ${\bar{\theta} \to \bar{\theta} + \alpha}$ 
due to averaging over $\bar{\theta}$. This means that the global charge $N$ is still conserved. Its Noether expression $$N = \int d^3 \mbs{x} \, \frac{\delta \mathcal{S}_{\mathrm{eff}}}{\delta \prt_t \bar{\theta}}\,,$$ however, includes a correction to Eq.~\eqref{N_charge} because $\mathcal{S}^{(2)}_{\mathrm{eff}}$ depends on $\partial_t \bar{\theta} = \omega$.

After finding $\mathcal{S}^{(2)}_{\mathrm{eff}}$, we again allow $\varphi_0$ and $\mu$ to vary with~$I \to \bar{I}$ via Eqs.~\eqref{phi0_choose} and \eqref{mu_0_def}. This time, the effective action is sensitive to $\vf_0$ at the weaker level~$O(\veps^3)$. 

The action~\eqref{field_eff_action}, \eqref{S_corr} allows us to compute corrections to the oscillon profiles {$\psi(r) = \psi^{(1)}(r) + \psi^{(2)}(r)$}, their fields $\varphi(\bar{I} + \delta I, \bar{\theta} + \delta \theta)$, charges, and energies. 

\begin{figure}
	\centering
	\includegraphics{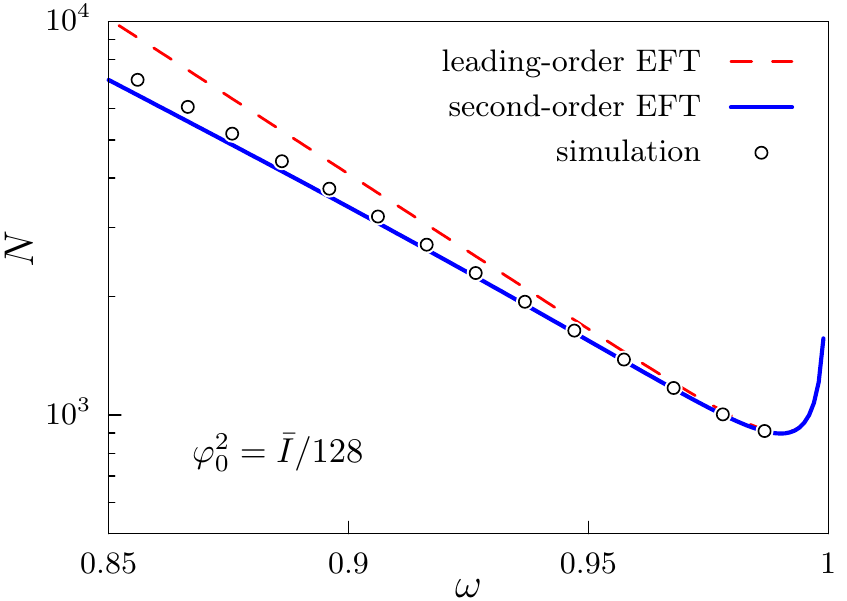}
	\caption{Comparison of the first- and second-order EFT predictions (lines) with exact simulations (circles). The plot shows the charge $N(\omega)$ of the monodromy oscillons at~${p=0.8}$. We intentionally ruined the precision of the theoretical calculation by detuning the parameter $\vf_0$.}
	\label{fig:Om_N_Corrs}
\end{figure}

Let us juxtapose the improved theory with exact simulations. However, we saw that the accuracy of our leading-order EFT is already comparable to the numerical precision. To see the progress, we intentionally spoil the theory by choosing a counter-intuitive auxiliary scale $\vf^2_0 = \bar{I}/128 \ll \bar{I}$, cf.\ Eq.~\eqref{lin_I_theta}. Besides, we use essentially non-quadratic potential with $p = 0.8$ and $\veps = 0.2$. Together, these deteriorations move the leading-order predictions for $N(\omega)$ away from the exact result, cf.\ the dashed line with the circles in Fig.~\ref{fig:Om_N_Corrs}. But the second-order EFT (solid line) is less susceptible to the impairment and agrees with the simulations. This demonstrates that higher-order corrections are capable of improving the effective theory, although they are impractical in the model under consideration.

It is worth noting that corrections to the effective action of even higher orders can be computed similarly to Eq.~\eqref{S_corr}, in two steps. First, solve Eqs.~\eqref{delta_eq_w_sources} to the required order in $\delta I$ and $\delta \theta$ and substitute the result into the action~\eqref{field_action} expanded to the same order. Second, average the resulting Lagrangian over $\bar{\theta}$. The possibility of performing calculations to arbitrary order exposes our effective theory as asymptotic expansion and clarifies its region of applicability.

\section{Discussion}
\label{sec:Discussion}
We developed a simple and precise analytic description of oscillons in scalar models with nearly quadratic potentials. The two cornerstones of our method are the correct choice of variables~$I,\,\theta$ in Eq.~\eqref{lin_I_theta} and the ``running'' (field-dependent) mass $\mu(I)$ in Eqs.~\eqref{V_lin+pert}, \eqref{mu_0_def}, and \eqref{phi0_choose}. We demonstrated that the effective action~\eqref{field_eff_action}, \eqref{S_corr} for $I$ and $\theta$ has the form of a systematic asymptotic expansion in the spatial derivatives and nonlinearities of the potential. Oscillons appear in this effective theory as non-topological solitons minimizing the energy at a given value of the adiabatic invariant~$N$. 

\begin{figure}[h]
	\centering
	\unitlength=1mm
	\begin{picture}(86,61)
		\put(0,0){\includegraphics{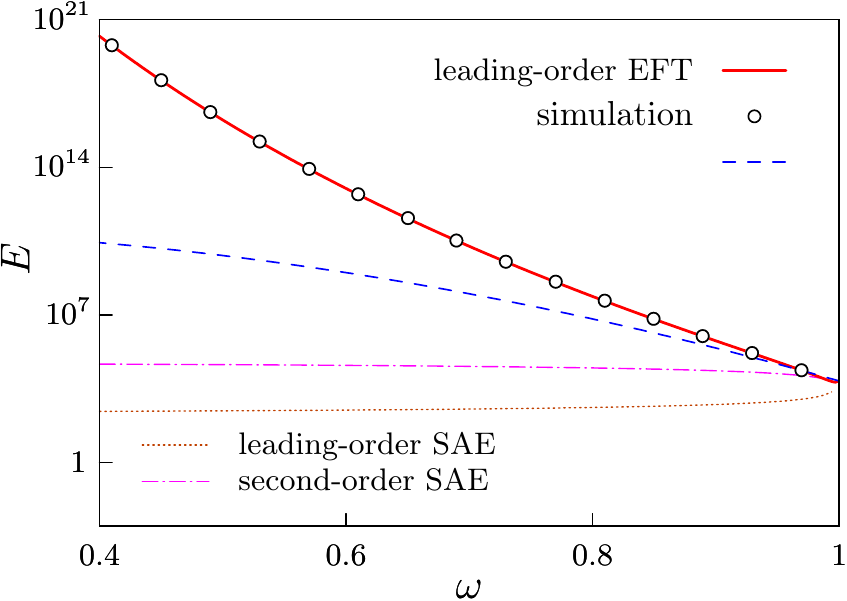}}
		\put(58.8,43.45){Eq.~\eqref{except_pot}}
	\end{picture}
	\caption{The energy $E(\omega)$ of three-dimensional monodromy oscillons at $p=0.95$ and $\veps = 0.05$.}
	\label{fig:Om_E_compare}
\end{figure}

The best part of our method is the possibility of computing corrections by keeping more terms in the expansion. At the same time, suitable choice of the ``running mass'' $\mu$ radically improves precision, making credible even the leading-order results. This trick with $\mu$ is inspired by the renormalization theory: it does the same job for the effective classical action as scale-dependent coupling constants do for the perturbative QFT. We demonstrate its power once again in Fig.~\ref{fig:Om_E_compare} by plotting the energy~$E(\omega)$ of oscillons in the monodromy model~\eqref{monodromy_welcome} as a function of their frequency~$\omega$ at~$p = 0.95$. Relative deviation of our leading-order theoretical prediction (solid line) from the exact simulation (circles) never exceeds $\Delta E / E \lesssim 0.06$  despite the fact that fields inside the oscillons are exceptionally strong at small $\omega$.

It is instructive to compare our new theory with other analytic approaches.
In the perturbative (small-amplitude) method, one expands the monodromy potential~\eqref{monodromy_welcome} in powers of $\vf$,
\begin{equation}
	\label{V_SAE}
	V = \varphi^2/2 - \veps \, \varphi^4/4 + \veps(1+\veps) \; \vf^6/12 + \ldots
\end{equation}
and then solves the field equation order-by-order in it~\cite{Kosevich1975, Dashen:1975hd, Segur:1987mg, Fodor:2008es,
  Fodor:2019ftc}. Generally speaking, such small-amplitude
  expansion (SAE) is valid only at $\omega \sim m=1$, since it can
be recast as series in the binding energy $(1 - \omega) \propto \vf^2$ of particles inside the oscillons~\cite{Kosevich1975}. This is apparent in Fig.~\ref{fig:Om_E_compare}, where the dotted and dash-dotted lines show the first two orders of SAE (two and three terms in Eq.~\eqref{V_SAE}, respectively). In the most interesting and wide frequency region $\omega \lesssim 0.9$ they considerably deviate from the simulations.

Another method employs expansion in~${\veps \equiv 1-p}$ characterizing nonlinearity of the monodromy potential~\eqref{monodromy_welcome}. At the first order and $|\vf| \gg 1$ one obtains~\cite{Olle:2020qqy},
	\begin{equation}
		\label{except_pot}
	V = \frac{\vf^2}{2} \left[1 + \veps  - \veps  \ln\vf^2  +  O(\vf^{-2}) + O(\veps^2 \ln^2 \!|\vf|)\right]. \!\!
\end{equation}
        This truncated model has a family of exactly periodic solutions with Gaussian spatial profiles~\cite{Dvali:2002fi,Kawasaki:2015vga, Olle:2020qqy} that approximate the monodromy oscillons~--- see the dashed line in Fig.~\ref{fig:Om_E_compare} showing their energies. We observe that Eq.~\eqref{except_pot} does a better job than the small-amplitude expansion, but fails at small $\omega$ when oscillon fields become exceptionally large,~$\veps \ln\! |\vf| \gtrsim 1$.

\begin{figure}
        	\centering
        	\includegraphics{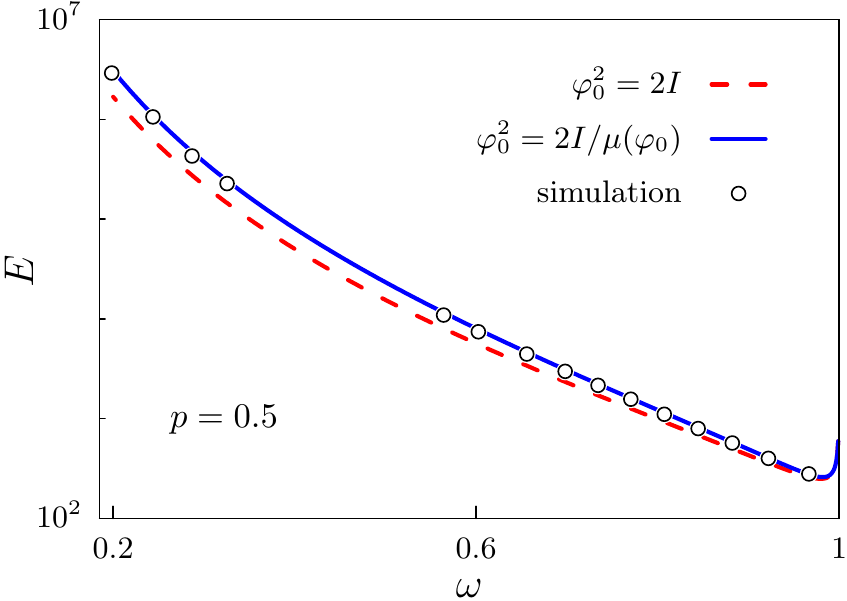}
        	\caption{Oscillon energies $E(\omega)$ in the case of
                    strongly nonlinear monodromy potential with~${p
                      = 0.5}$ (points) compared to the two leading-order
                    EFT predictions (solid and dashed lines). The latter lines differ by the 
                    choice of the EFT scale~$\varphi_0$.}
        	\label{fig:0.5p}
        \end{figure}

        In contrast, our effective theory universally applies and
        remains precise at all frequencies and oscillon
          amplitudes. Its  relative accuracy 
          is rather controlled by the anharmonicity parameter $\varepsilon
          \equiv 1-p$ of the potential. For example, in the 
          model of monodromy inflation~\cite{Lozanov:2014zfa, Liu:2017hua,
            Cotner:2019ykd, Lozanov:2023aez} with~${p = 0.5}$ the
          leading-order EFT results for the oscillon energies are
          offset by $\Delta E / E \sim 0.4$ from the exact data,
          cf.\ the dashed line and the  circles in
          Fig.~\ref{fig:0.5p}. One can make a better 
          choice of the EFT parameter~$\varphi_0$, however: find it by
          solving the equation $\varphi_0^2 = 2I /
          \mu(\varphi_0)$. Then the relative 
           error drops down to
          $\Delta E/E \sim 0.1$ which is surprisingly small; see the
          solid line in Fig.~\ref{fig:0.5p} and 
          Footnote~\ref{fn:smart_mu}  for details. This suggests that
          wise choice of the EFT scale is capable of curing the method
          even in the case of significantly nonlinear potentials\footnote{This does not
          mean, however, that the EFT series converge well in this
          case, since the expansion parameter~$\varepsilon=0.5$ is
          large.}. 
        
        We anticipate that the results of this paper will be helpful
          for analytic calculations in oscillon cosmology. In
          particular, theoretical relations between oscillon amplitudes, energies,
          frequencies, and adiabatic charges 
          are important whenever these objects represent dark
          matter~\cite{Olle:2019kbo, Olle:2020qqy} or generate 
          gravitational waves after inflation~\cite{Amin:2011hj,
            Zhou:2013tsa, Sang:2019ndv}. Even more results can be
          obtained by applying our method in other models with nearly
          quadratic potentials akin to
          Eq.~(\ref{monodromy_welcome}). 

        But the most interesting development 
          would be to adopt our approach for 
            calculation of the oscillon evaporation
          rates~$\Gamma$. Segur and Kruskal evaluated them
        analytically in the framework of small-amplitude
        expansion~\cite{Segur:1987mg}, see
        also~\cite{Fodor:2008du,Fodor:2009kf,Fodor:2019ftc}. Namely, 
        they demonstrated that $\Gamma \propto
        \exp(-\mathrm{const}/g_0)$ is nonperturbative in the expansion
        parameter $g_0 \propto \vf$. Our technique is different, but
        it also has the form of asymptotic expansion. Its formal
        parameter $g$ can be introduced in Eq.~\eqref{V_lin+pert}
        as $$V = \mu^2 \vf^2 / 2 + g^2\, \delta V,$$ where $g=1$ is
        the physical value. 
Once this is done, the rescaling $\mbs{x} = \mbs{\tilde{x}}/g$ brings
$g^2$ in front of the second small term with spatial derivatives. It
would be fascinating to apply methods of nonperturbative resummation
in Refs.~\cite{Segur:1987mg,Fodor:2019ftc, Fodor:2008du,Fodor:2009kf}
to our series, thus getting a general expression for $\Gamma$ in
models with nearly quadratic potentials. The latter calculation,
however, lies beyond the scope of the present paper.

\begin{acknowledgments}
We thank E.~Nugaev and A.~Panin for fierce discussions. This work was supported by the grant RSF \mbox{22-12-00215}. Numerical calculations were performed on the 
Computational Cluster of the Theoretical Division of~INR RAS.
\end{acknowledgments}

\bibliography{monodromy}{}
\end{document}